\pdfoutput=1 
\documentclass[manuscript, nonacm]{acmart}
\usepackage[utf8]{inputenc}
\usepackage{subfiles}
\usepackage{tikz}
\usetikzlibrary{positioning}
\usepackage{url}
\title[Justice in Misinformation Detection Systems]{Justice in Misinformation Detection Systems: An Analysis of Algorithms, Stakeholders, and Potential Harms}

\author{Terrence Neumann}
\affiliation{%
\institution{University of Texas at Austin}
\streetaddress{2110 Speedway}
\city{Austin}
\state{TX}
\country{USA}
\postcode{78705}}
\email{Terrence.Neumann@mccombs.utexas.edu}

\author{Maria De-Arteaga}
\affiliation{%
\institution{University of Texas at Austin}
\streetaddress{2110 Speedway}
\city{Austin}
\state{TX}
\country{USA}
\postcode{78705}}
\email{dearteaga@mccombs.utexas.edu}

\author{Sina Fazelpour}
\affiliation{%
\institution{Northeastern University}
\streetaddress{360 Huntington Ave}
\city{Boston}
\state{MA}
\country{USA}
\postcode{02115}}
\email{s.fazel-pour@northeastern.edu}

\acmConference[FAccT 2022]{ACM Conference on Fairness, Accountability, and Transparency}{June 21--24}{Seoul, South Korea}
\acmBooktitle{FAccT 2022: ACM Conference on Fairness, Accountability, and Transparency 2022}

\setcopyright{acmcopyright}

\begin{abstract}

Faced with the scale and surge of misinformation on social media, many platforms and fact-checking organizations have turned to algorithms for automating key parts of misinformation detection pipelines. While offering a promising solution to the challenge of scale, the ethical and societal risks associated with algorithmic misinformation detection are not well-understood. In this paper, we employ and extend upon the notion of \emph{informational justice} to develop a framework for explicating issues of justice relating to representation, participation, distribution of benefits and burdens, and credibility in the misinformation detection pipeline. Drawing on the framework: (1) we show how injustices materialize for stakeholders across three algorithmic stages in the pipeline; (2) we suggest empirical measures for assessing these injustices; and (3) we identify potential sources of these harms. This framework should help researchers, policymakers, and practitioners reason about potential harms or risks associated with these algorithms and provide conceptual guidance for the design of algorithmic fairness audits in this domain.

\end{abstract}

\begin{document}


\maketitle

\section{Introduction} 
Misinformation has unfortunately become ubiquitous in our day-to-day life. The surge in the spread of misinformation on social networks has raised deep concerns regarding the well-being of individuals, groups, and democratic societies. Faced with the scale of the challenge, many platforms have turned to algorithms to automate content moderation decisions, and fact-checking organizations are increasingly adopting algorithmic tools for support. Algorithmic content moderation appears to be widely adopted by some platforms. For instance, Facebook reportedly labeled 180 million messages as potentially misleading during the 2020 election season, while Twitter says it labeled 300,000 during the same period~\cite{lerman_kelly_2020}. Further, Facebook claims to have taken 1.8 billion actions against fake accounts violating internal policies from July to September, 2021~\cite{transparency_center_2021}.

However, there is a lack of attention to the ethical risks that algorithmic biases, similar to those found in systems used to assist with hiring, policing, and medical diagnostic decisions, may present in automated misinformation detection systems as well. Specifically, there has been no systematic study of the senses and sources of injustice that can plague automated misinformation detection systems. This gap is problematic not only because of what is at stake in how we devise and evaluate sociotechnical responses to misinformation; it is troublesome particularly because the much-discussed frameworks of justice that are suitable for allocation settings are not sufficient to capture all dimensions of justice in information ecosystems.

In this paper, we develop a framework for characterizing, detecting, and diagnosing the sources of injustices that can affect the performance of large-scale automated misinformation detection systems. We begin with discussing relevant existing work work in Section~\ref{sec:rel}. In Section~\ref{sec:tax}, we introduce our framework by building on the notion of \textit{informational justice}. Developed by \citet{mathiesen2015informational}, this multifaceted notion identifies three key dimensions of justice---participatory, recognitional, and distributive---that are pertinent to three types of stakeholders implicated in informational items---sources, subjects, and seekers of information, respectively. We extend this framework, to capture the additional stakeholders involved in misinformation detection pipelines and the further senses of justice that corresponds to them. In Section~\ref{sec:situating}, we put this framework to use by situating it in the misinformation detection pipeline. Focusing on each stage of the pipeline, we explain why injustices of a specific sense might emerge in that stage, how we might empirically track their presence and trace their potential sources. 

Throughout, we show how the framework provides practitioners, policy-makers, and platform executives with a robust tool for investigating the complex issues of justice, involving a myriad of diverse stakeholders, associated with automatic content moderation.

\section{Background and Related Work} 
\label{sec:rel}

\subsection{Machine Learning and Algorithmic Injustices}
\label{subsec:fairML}

The study of injustices that may result from the deployment of machine learning systems has largely focused on the risks of algorithmic bias~\citep{barocas2016big}, paying special attention to classification tasks that may inform the allocations of benefits and burdens~\citep{fazelpour2021algorithmic, mitchell2021algorithmic}. In such contexts, researchers have emphasized the risks of compounding inequalities~\citep{eubanks2018automating} and historical injustices~\citep{hellman2021big} in domains such as criminal justice~\citep{angwin2016machine}, human resources~\citep{de2019bias}, and healthcare~\citep{obermeyer2019dissecting}. While these relate primarily to allocative harms~\citep{barocas2017problem}, representational harms have also received attention in the information retrieval context, with special attention devoted to stereotyping~\citep{blodgett2020language, mitchell2020diversity}. 

Most closely related to our work is research on algorithmic fairness in the context of hate speech detection, as it constitutes another type of harmful online content~\citep{banko2020unified}. Research in this space has empirically shown that automated hate speech detection systems may incorrectly label African American English tweets as toxic at disproportionate rates~\citep{sap2019risk}, and similar patterns of errors are observed for statements discussing identity of belonging to minoritized groups, such as posts saying ``I am gay"~\citep{dixon2018measuring}. This line of research has also provided critical perspectives of the datasets used to train these systems~\citep{madukwe2020data,vidgen2020directions}, and conducted ethnographic studies around expectations and responses to these systems~\citep{laaksonen2020datafication}. 

\subsection{Governance of Content Moderation}
Governance mechanisms regarding content moderation vary around the world \cite{gillespie_2021}. For instance, in the United States, Section 230 of the Communications Decency Act of 1996 remains a crucial piece of legislation impacting content moderation. The legislation intended to reform tort law in light of the widespread impacts of the internet \cite{sylvain2018discriminatory}. In a world becoming increasingly digitally connected, owners of digital platforms that connected massive amounts of internet users - either for purely social reasons or for commercial exchanges - were afraid that they might face an untenable burden of liability for the behaviors of their user base. In response to these fears, Section 230 effectively removes liability for these internet companies - called ``intermediaries'' - in all cases except those in which they ``materially contribute'' to this harmful user content.\footnote{Jones v. Dirty World Entertainment Recordings LLC, 755 F.3d 398 (6th Cir. 2014)} While this doctrine allowed online media and commerce to flourish without the constant threat of costly law suits, it also fostered an environment that was, at best, neutral to those harmed by user generated content. By completely detaching these companies from the harms stemming from their users' behaviors, a poor system of incentives developed in which genuinely harmful content proliferated. 

Social media companies have, to date, used a largely tactical and supposedly neutral approach to content moderation, characterized by an ever-growing list of ad-hoc rules formed in response to political pressure from lawmakers or interest groups. This, according to \citet{information_quality}, is unjust and ``creates an uncertain epistemic environment ... inviting claims of bias, favoritism, and censorship. [p. 8]'' As an example of the limited use and potential harms of such ad-hoc tactics, consider Facebook's response to the claim, especially made by the politically conservative, that misinformation labels displayed by social media companies are politically biased. In response to this, Facebook is reported to have changed some of its content standards to allow content that previously would have been flagged as misinformation to remain unlabeled and unaltered~\cite{wapo_2021}. That is, the response consisted of altering content moderation standards \textit{as they relate to only some sources of information} to mitigate the claim. Yet, this response could lead to potential discrepancies in the quality of labeled misinformation for different seekers of information, and the harms associated with this will be discussed in subsequent sections.

\subsection{Harms of Misinformation}

Many significant harms associated with misinformation and disinformation have been documented. For instance, researchers have observed harms related to: the adoption of addictive habits ~\cite{misinformation_harms_e_cigs}; receiving misleading health advice during a pandemic \cite{misinformation_harms_virus}; democracy and social institutions \cite{misinformation_harms_democracy}; beliefs about climate change \cite{misinformation_harms_climate}; and situational awareness in humanitarian crises \citep{tran2021investigation}. 

Search engines and social media platforms may utilize targeted advertising as a source of revenue, in which content creators can promote their content to users with certain demographic or personality characteristics that yields a higher propensity to engage with the content \cite{targeted_ad1, knoll_advertising_2016}. Research on "computational propaganda" has highlighted how adversaries intent on causing societal strife have used targeted advertising to show ``weaponized'' information likely to manipulate specific groups of users~\cite{computational_propaganda, nadler2018weaponizing}. (Some have also called this domain of research ``social cyber-security'' \cite{beskow2019social}.) Importantly, in some cases such attacks are facilitated by the automation of key tasks by social media platforms. For instance, there is evidence that algorithms implemented by Facebook created categories that clustered users based on their propensity to engage with anti-Semitic articles, and this information was made accessible to those willing to pay for it~\cite{antisemitic}.

In addition to being a potential risk to democracy and national security, targeted advertising can also pose an \textit{epistemic} threat to communities. This can happen, for instance, when the technology facilitates the promotion of fraudulent products and conspiracy theories. Reports regarding ads on Facebook for a hat that could ``protect [one's] head from 5G cell-phone radiation'' are an instance of this threat, particularly when they coincide with the increased popularity and circulation of the related conspiracy theory on social media~\cite{targeted_ad2}. Further investigation of such ads point to the role of target advertising, such as when Facebook determines that a user is interested in the product category ``pseudoscience''~\cite{targeted_ad2}. 

Our work is crucially different from these lines of research in that we do not seek to study the harms associated to misinformation itself, but rather the injustices that particular stakeholders may encounter when deploying algorithmic tools to combat this phenomenon.

\subsection{Algorithmic Misinformation Detection}

To combat the scale of misinformation, large social media companies (like Facebook) have developed algorithms and policies to identify and act upon misleading information shared on their platform. Facebook's system, in particular, automatically detects factual claims worth checking. These claims would then be sent to an independent fact-checker to assess its veracity. Facebook also employs algorithms to determine if claims have been previously fact-checked by the human fact-checkers. If the result of this assessment is that a claim is false or misleading, Facebook ``add(s) warnings and more context to content rated by third-party fact-checkers, reduc(es) their distribution, and remov(es) misinformation that may contribute to imminent harm'' \cite{meta_ai}. 

Algorithmic misinformation detection can be composed of many sub-tasks, which some systems tackle independently while others attempt to solve in an end-to-end fashion. While the specifics of these tasks may evolve and change over time, we draw from~\citet{survey_AFC} to differentiate between three core (sequential) tasks: (1) \emph{Check-worthiness}, which aims to spot factual claims that are worthy of fact-checking~\cite{checkthat, automated_claim_detection, claimbuster, twitter_metadata}, (2) \emph{Evidence retrieval} of potential evidence for identified claims~\cite{information_retrieval_memory_network,evidence_aware_model,thai_fake_news,FAKTA,politifact_evidence,wiki_evidence,lee_2018_improving} , and (3) \emph{verdict prediction}, which aims to establish the veracity of a claim~\cite{evidence_aware_model,supervised_fake_news_detection,FAKTA}. In a survey on the topic by \citet{survey_fake_news_detection}, the authors identify how misinformation can be detected from four perspectives: (1) the false knowledge it carries; (2) its writing style; (3) its propagation patterns; and (4) the credibility of its source. In subsequent sections, we consider ``general'' machine learning methods with features that account for these four perspectives related to fact-checking.

\section{Justice \& Stakeholders in Misinformation Detection}
\label{sec:tax}
Misinformation detection pipelines consist of varied and many tasks, each of which can give rise to many ethical concerns. Precisely articulating these concerns requires a coherent normative framework for identifying the relevant stakeholders, mapping their legitimate rights and interests, and disentangling the distinct claims that they may justifiably have to justice. Therefore, well-known frameworks that focus on only a single stakeholder (i.e. the source of information) and single ethical notion (free speech) will be inadequate to describe the complex ethics of the information environment. 

In this section, we provide this framework by building on Mathiesen's \cite{mathiesen2015informational} notion of \textit{informational justice}---a multidimensional concept characterized in terms of ``justice for persons and communities in their activities as seekers, sources, and subjects of information''~\cite[p. 199]{mathiesen2015informational}. Mathiesen's \cite{mathiesen2015informational} framework originated in the context of information sciences, and so most directly applies to discussions of justice surrounding \textit{(mis)information}. In order to comprehensively capture the concerns regarding misinformation \textit{detection}, therefore, we need to appropriately extend the framework. In this section, we explain both the framework as well as our extensions. To see why a multidimensional notion of justice is needed in the first place, we begin by contrasting the situation in misinformation detection with the more familiar discussions of justice in allocation.

\begin{figure}
 \centering
    \includegraphics[width=0.6\textwidth]{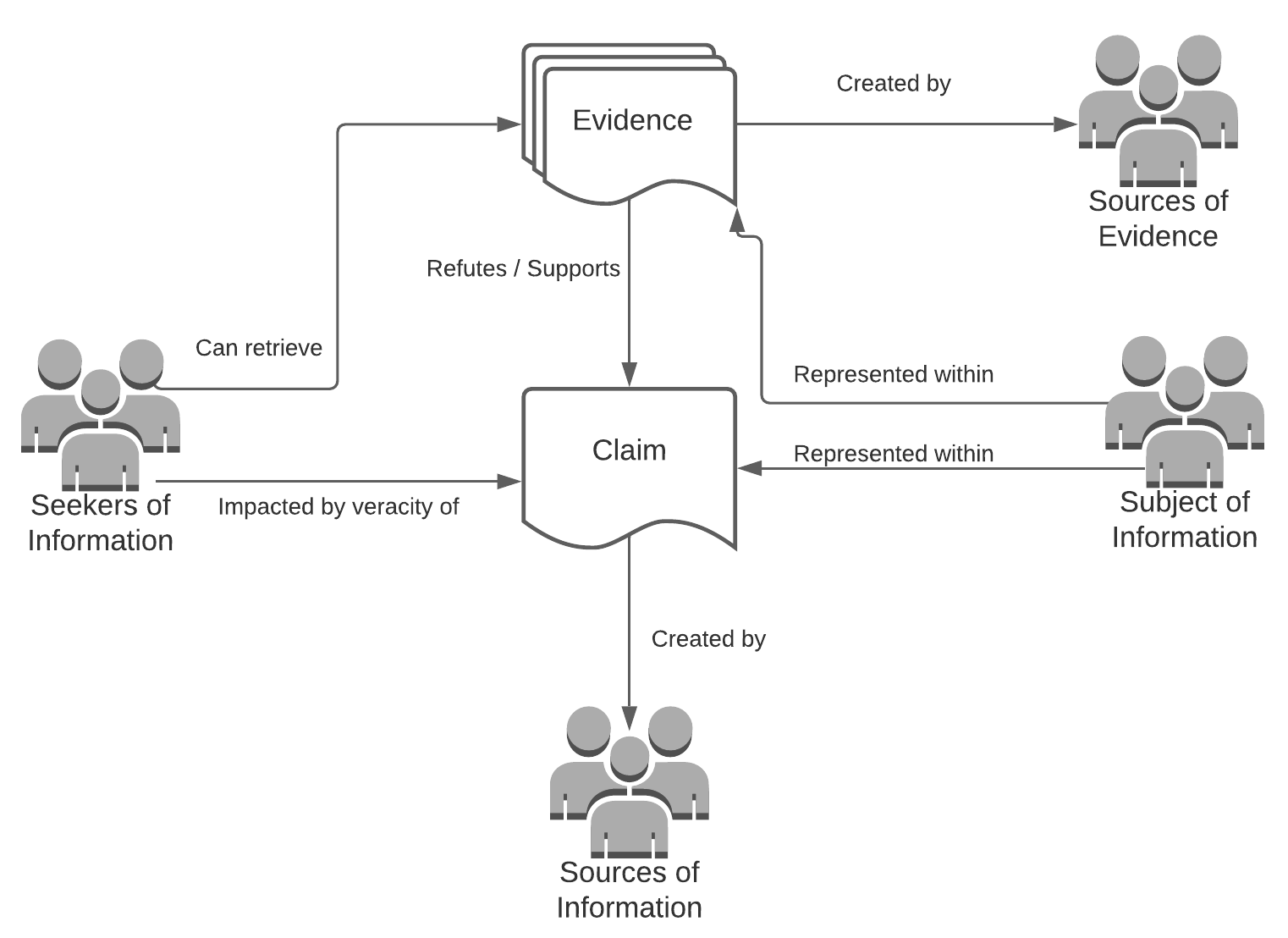}
    \caption{This figure demonstrates how stakeholders are related to two informational items (claims and evidence) in automated misinformation detection, borrowing heavily from the informational justice framework posited by \citet{mathiesen2015informational}.}
    \label{fig:1}
\end{figure}

In common discussions of algorithmic fairness, each data point typically pertains to individuals (e.g., a job candidate in the case of hiring, a school district in the case of public funding allocations). These individuals tend to occupy a single direct stakeholder role, namely, that of decision subject. In allocation situations, these decision subjects encounter a decision-maker (e.g., an employer, a public funding body) who must decide how to distribute scarce resources and opportunities among them. Generally, here it is the decision subjects who have an invested interest in justice---more specifically distributive justice~\cite{elster1992local}---, while the primary concerns of the decision-makers (to whom the demands of justice apply) might be in values such as efficiency (e.g., for profit or for social welfare).\footnote{The decision-makers might also be interested in being just, and, when we go beyond direct stakeholders, particular communities and groups as well as general public might also prize distributive justice.} 

In the case of informational items, in contrast, each data point typically relates to \textit{multiple} stakeholders, who interact with or are directly implicated in the informational item. A tweet or a book, for example, has an \textit{author} and an \textit{audience}, and it can have a particular individual or group as its \textit{subject matter}. In Mathiesen's informational justice framework, these individuals or groups map onto three stakeholder roles: \textit{seekers of information}, \textit{sources of information}, and \textit{subjects of information}. In the case of misinformation detection, in addition to the informational item that is the focus of evaluation (e.g., a tweet about which there has been a complaint), there is an additional type of informational item: \textit{evidence} (e.g., in the form of the original complaint, or subsequent evidence gathered in verification). We identify an additional stakeholder role that is relevant to the evidence gathered during various stages of the process---from check-worthiness to truth verification): \textit{sources of evidence}. Below, we describe each of these stakeholder types in more detail and explain the notion of justice most directly associated with them. Figure~\ref{fig:1} maps out these stakeholders in relation to the two informational items involved in misinformation identification: a claim and evidence regarding the claim. 

\paragraph{Sources of Information and Participatory Justice.}
Sources of information in misinformation detection are authors of articles or content creators that produced or else posted a claim for dissemination over social media. Conceptions of justice relating to the sources of information involve balancing the right to participate in the ``marketplace of ideas'' with the duty to act ``appropriately'' (often as dictated by either federated or centralized user agreements) and to not intentionally deceive others. 
The notion of justice relevant to the sources of information thus pertains to participation. \textit{Participatory} justice entails that ``all members of society should have opportunities to communicate their point of view alone or in concert with others, to have that point of view taken into account, and to take part in shared decision making about the provision of information resources.'' In the context of misinformation detection, this entails certain rights of all users to have their opinions or thoughts heard by those connected to them, and, when applicable, to a broader audience. This would imply that all sources of information should be treated equally in a given social media ecosystem, regardless of their affiliation with any social-cultural group.

\paragraph{Subjects of Information and Recognitional Justice.}
Subjects of information are represented within claims made by sources of information. Subjects can be individuals or groups, or they can be conceptual topics. When subjects are individuals or groups, a just system would ensure that their reputation faces no undeserved harm from inaccurate claims made by others. More broadly, \textit{recognitional} justice implies that ``contents of information available within the information environment should include fair and accurate representations of all members of society.'' In the current context, recognitional justice has particular importance, for example, in cases of false claims and derogatory language directed at particular social-cultural groups.

\paragraph{Seekers of Information and Distributive Justice}

Seekers of information are people or groups that may be impacted by the presence of information about a topic that is of relevance to them and the veracity of the claims (tweet, article, images, etc) about that topic circulating online. Given the foundational ways it shapes and interacts with our resources and opportunities, information can be seen as a critical good in our societies. Accordingly, the notion of justice that is particularly pertinent in relation to seekers of information is largely \textit{distributive} in nature. That is, justice here pertains to how informational resources are distributed among individuals and groups, and it is undermined by disparities in the availability or quality of relevant information. On social media, for example, such disparities can happen when users from certain demographic backgrounds do not see potentially relevant information (e.g., do not see a relevant job ad because of a clustering of users in homophilic networks) or receive lower quality information (e.g., receive ads about irrelevant jobs perhaps due bias in the recommendation system). Similar issues apply in the case of misinformation detection. Distributive injustice can happen, for example, when misinformation targeted at a marginalized community is more likely to circulate unhindered. Like other cases of distributive justice, what precisely the demands of justice consist in here requires specifying (and agreeing upon) the relevant resource that is being distributed (e.g., information, ``quality'' of information in some sense) as well as the appropriate rule for the distribution of that resource.\footnote{The appropriateness of the rule might depend on the context.}

\paragraph{Sources of Evidence and Epistemic Injustice}
Individuals generating the second informational item---that is, evidence---constitute an additional stakeholder. Depending on the system design and the stage of the misinformation detection pipeline, this stakeholder may be a user of a platform who flags a claim (e.g., a twitter post), a fact-checker who independently analyzes the claim, a crowd worker surveyed to assess the veracity of the claim, or the author of an article that provides a stance towards the claim. In each case, the activities of these individuals results in the generation of a piece of evidence about the focal claim. Insofar as their activities directly shapes the information ecosystem, \textit{sources of evidence} also may have a claim to participatory justice. 

Here, however, we want to focus on another source of (in)justice that applies to sources of evidence more specifically, namely, \textit{epistemic justice}. To the extent that misinformation detection systems fail to incorporate or else disregard the relevant knowledge produced by some sources of evidence, they may result in epistemic injustice. Particularly relevant in the misinformation detection context might be cases of \textit{testimonial} injustice, where the claims of a source of evidence (e.g., that a tweet constitutes hate speech) receives an unfair deficit of credibility downstream (from other sources of evidence or an algorithm), and diminishes the source's capacity as a \textit{knower}~\cite{fricker2007epistemic}. This type of epistemic injustice can be particularly problematic if it systematically affects members of marginalized communities.

Epistemic justice is not the only type of justice that is relevant to sources of evidence. Depending on a source of evidence's role, considerations of distributive justice can also come into play. Consider, for example, the emotional toll of hate speech on individuals from the targeted group. If there is a disparity in the amount and circulation of such speech such that individuals from affected groups are more likely to be forced to report and flag these items (that is, to voluntarily assume the role of sources of evidence), to that extent the information platform exhibits distributive injustice, potentially compounding existing disparities. Similar points apply to the toll of such speech on the mental health of crowdsource workers who are tasked to label or else assess the veracity of the claim.

In this section, we primarily focused on notions of justice in relation to \textit{claims} (or evidence pertaining to them) that form the input to misinformation detection systems. In addition to this, issues of distributive justice become salient in relation to the \textit{misinformation detection systems} themselves, insofar as the performance and impacts of these systems shapes how the benefits (e.g. information quality control) and burdens (e.g. over-scrutiny) of the technology are distributed. For example, as highlighted in~\ref{subsec:fairML}, in many instances the classification errors of existing algorithms disproportionately harm members of disadvantaged groups, thus resulting in concerns about distributive injustice.

\section{Situating Algorithmic Injustice in the Misinformation Detection Pipeline}
\label{sec:situating}
\begin{figure}
 \centering
    \includegraphics[width=\textwidth]{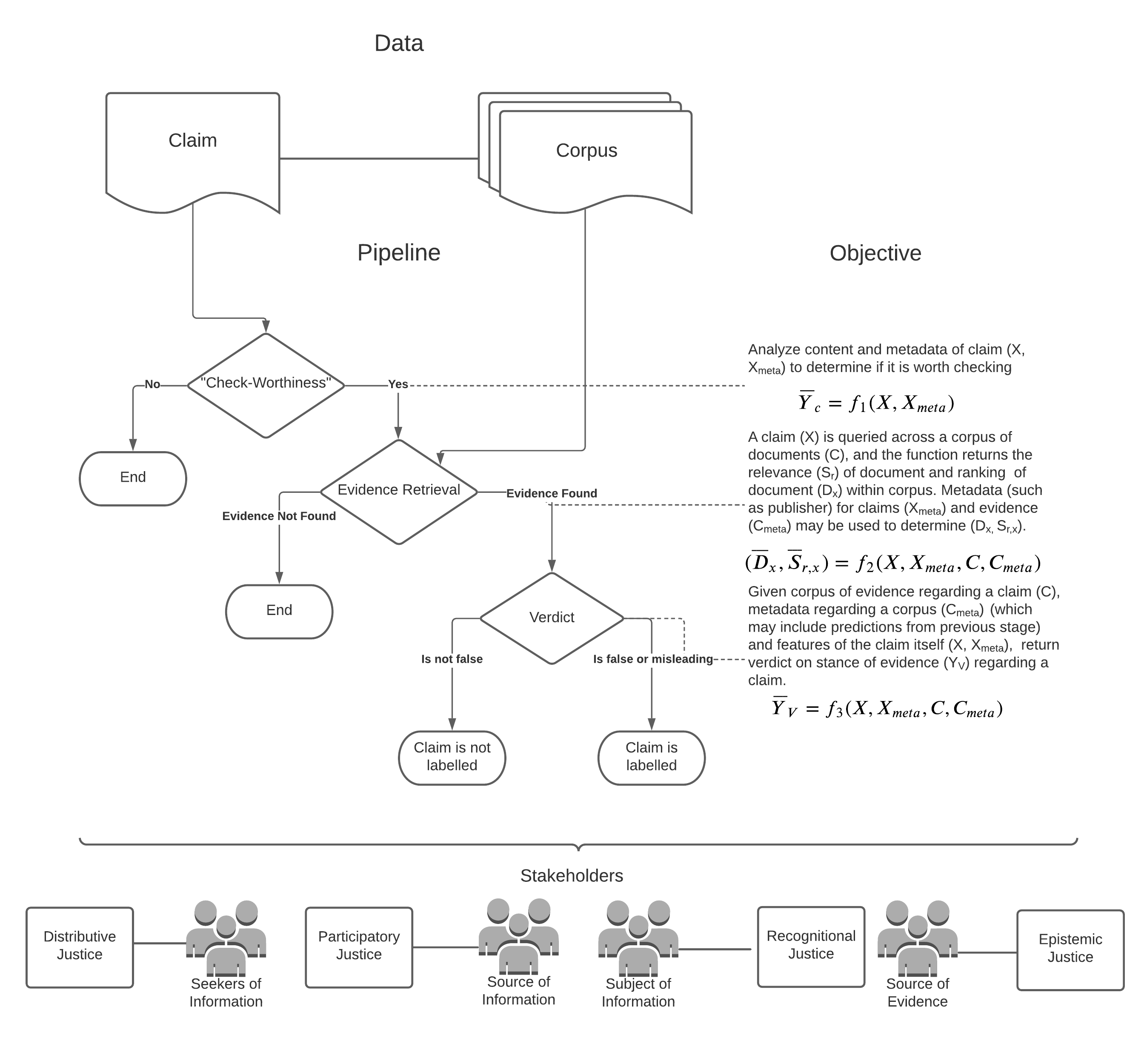}
    \caption{A generic framework, used in conjunction with Figure \ref{fig:1}, to examine issues of justice regarding algorithmic decision-making in the automated misinformation detection pipeline.}
    \label{fig:2}
\end{figure}

Machine learning can be employed to automate the entire misinformation detection pipeline, or to assist in specific stages of it. Figure \ref{fig:2} identifies the three central stages of the process, and provides a generalized view of the target objective at each of them. While specific design choices may vary across systems (e.g., choosing if check-worthiness is a multi-class label prediction problem or a regression problem), this general formulation enables us to situate the different types of injustices that may emerge at distinct functional stages of these systems. 

Studying, anticipating or mitigating algorithmic harms requires the identification of stakeholder(s) affected by incorrect algorithmic predictions, together with the notion(s) of justice relevant to a stakeholder when a certain type of error is made. For example, some errors might result in undue scrutiny of speech or disproportionate harms stemming from the spread of misinformation, depending on whether the algorithm is over- or under-scrutinizing the content relevant to a particular group. In this section, we discuss the potential harms associated with different algorithms in automated misinformation detection, grounded on the taxonomy introduced in Section~\ref{sec:tax}. For each algorithmic stage in the pipeline, we discuss: (1) a generalized framework/objective for the machine learning task; (2) the relevant sense of injustice to different stakeholders when affected by errors in algorithmic decisions, and (3) an overview of potential sources of these injustices. 

\subsection{Automated Claim Detection}

\subsubsection{The Machine Learning Task}

The first task in the automated misinformation detection pipeline is identifying ``check-worthy'' claims. According to \citet{claimbuster}, check-worthy claims are claims for which the ``general public'' would be interested in assessing their validity. For fact-checking organizations, these would be the claims (out of broader possible universe of claims) that they choose to evaluate. Snopes, for instance, says this about how they decide what to fact-check: ``We have long observed the principle that we write about whatever items the greatest number of readers are asking about or searching for at any given time, without any partisan considerations''~ \cite{snopes_2021}. It is worth noting that the motivations for identifying a subset of check-worthy claims may vary depending on the context and system design. Most commonly, check-worthiness is a prioritization mechanism to allocate the scarce resource that is human fact-checkers. Given the scalability of algorithmic fact-checking, such scarcity of fact-checking resources could be altered, but check-worthiness may remain as a relevant mechanism to screen content that should be fact-checked, conditioned on the nature of the claim. 

Several model architectures have been proposed for detecting check-worthiness of textual data (e.g.~\cite{checkthat, automated_claim_detection, claimbuster, twitter_metadata, checkworthy_multitask}). \citet{claimbuster}, for instance, gather numerous expert crowd-sourced labels for political speech sentences as either \emph{Non-Factual Sentence} (opinion, belief), \emph{Unimportant Factual Sentence} (facts that are not relevant to general pubic), and \emph{Check-Worthy Factual Sentence} (facts that are relevant to general public). They then extract sentence sentiment, named entities, and tf-idf features from the vocabulary and optimize a support vector machine classifier predicting check-worthiness. As common in misinformation detection, this system relies on training data from a very specific, and likely not generalizable, corpus: political speeches. 

CheckThat!, an annual misinformation prediction conference, has provided a dataset that contains 629 labeled English tweets pertaining to COVID-19 as a benchmark dataset~\cite{checkthat}. The researchers gathered survey responses to five questions that attempted to capture the check-worthiness of each tweet. Participants in the machine learning competition were then asked to aggregate these labels and develop a model to predict check-worthiness for novel tweets. 

Researchers investigating COVID-19 misinformation adopted a very similar approach to \citet{checkthat} for crowd-sourcing labels to detect claims made on social media, except instead of aggregating the label, they implemented a multi-task learning approach, in which the correlated survey questions used to determine check-worthiness were modelled jointly~\cite{checkworthy_multitask}.

Some researchers have also framed the task as rumor detection, and have found success using metadata on social media (author information, rate of decay of engagement, reactions, and replies) jointly with semantic representations of the content to spot early propagation of rumors that should be fact-checked \cite{twitter_metadata}. 

In general, we frame the machine learning task as trying to recover the function:

\begin{equation}
    \overline{Y}_{c} = f_{1}(X, X_{meta})
    \label{eq:1}
\end{equation}

Where \(\overline{Y}_{c}\) is the predicted check-worthiness label, score or vector of labels (binary, multi-class, or a tuple in the context of multi-task learning) for a given claim, \(X\) is a semantic representation of the claim, and \(X_{meta}\) is a vector of metadata associated with the claim (likes, reposts, etc). If a piece of content is predicted to be check-worthy, then it will pass to further stages in the misinformation detection pipeline, where evidence will be gathered in order to produce a veracity score or label. The check-worthiness algorithm is concerned with reducing the scale of information for which evidence retrieval is necessary.

\subsubsection{Issues of Justice}

Under-scrutiny of the claims reported by, of interest to, or about a particular group can lead to a distributive injustice for sources of evidence, seekers and subjects of information, as a key moderator of the informational good (fact-checking) is not being distributed evenly amongst groups. Seekers of information may be harmed because they receive a comparatively lower standard of information quality, allowing them to more directly feel the harms of misinformation than others. Recall the example from Section \ref{sec:rel}, where Facebook's seemingly ad-hoc content moderation policy led to claims of bias from conservative sources of information--a claim of participatory injustice in our framework. In response, Facebook changed the standards for evaluating misinformation from conservative sources of information, reducing scrutiny of some of these sources. By changing the standard for misinformation detection for specific groups (in this case, conservatives), harms fall upon (conservative) seekers of information, who are now more likely to be exposed to misinformation. Under-scrutiny errors at the claim detection stage may also affect subjects of information, who could be harmed if false claims were made against a community or an individual that could damage their reputation or fuel stigma and stereotypes and these claims were not prioritized to be fact-checked. Thus, failing to assess these claims could result in a recognitional injustice, by enabling inaccurate representations of members of society. Finally, in settings where user reports are an input to check-worthiness assessment, a disproportionate lack of attention to claims flagged by a community could constitute an epistemic injustice, which may additionally result in some of the injustices listed above for the same individuals and/or communities, when their role in relation to a claim is multi-faceted. While the type of injustice stemming from under-scrutiny will be different for different stakeholders, they all correspond to false negative errors. Thus, assessing \emph{equal opportunity}~\citep{hardt2016equality} as an algorithmic fairness metric is of relevance, where the sensitive attribute category will vary conditioned on the stakeholder. 

Over-scrutiny may also constitute harms, although these will depend on the interaction with other portions of the system. By itself, marking something to be verified does not necessarily harm sources, seekers or subjects, especially if we assume perfect accuracy in the fact-checking stage. However, that is an unrealistic assumption, and it is necessary to consider fairness under composition~\cite{dwork2018fairness}. In particular, it has been shown in the algorithmic fairness literature that making the same type of errors, e.g. false negatives, when the output of one algorithm determines the input of the following one will result in compounding imbalances, which may correspond to compounding injustices~\cite{de2019bias}.  Thus, monitoring \emph{false positive rates} associated with predicted check-worthiness is especially pressing in conjunction with the errors at other stages of the pipeline. 

\subsubsection{Sources of Harms}

Check-worthiness is an ambiguous task, and those designing a system must first decide how to define what being check-worthy means, and establish a quantitative approximation to it. This problem formulation stage may already induce bias. For example, check-worthiness may be defined as what interests the ``general public"~\cite{claimbuster}, or what ``the greatest number of readers are asking about" \cite{snopes_2021}. Clearly, a simple focus on majorities disregards minority communities along multiple axes. Misinformation targeting specific minorities, e.g. an immigrant community, will by definition not be considered check-worthy unless it also affects other groups, constituting a distributive injustice for this group of seekers of information. Similarly, when these communities participate in the ecosystem as sources of evidence, by reporting or searching for certain items, their input will be neglected unless it matches the interest of majority groups, resulting in an epistemic injustice. 

Thus, the harms associated with false negative errors, or under-scrutiny of claims, may be disproportionately concentrated in certain groups as a result of \textit{measurement bias} in the training labels~\citep{jacobs2021measurement}. Measurement bias--when an inappropriate proxy is used to measure a construct \cite{machine_learning_bias_general}--may result from a choice of metric, e.g. the most frequently searched items, or from ambiguity in the formulation of survey questions used to collect check-worthiness labels. For instance, questions inquiring whether something is of interest to the ``general public" may elicit very subjective responses regarding who the ``general public'' is, and may downplay claims that are highly relevant (with potentially high stakes) to a particular minority group. This issue may be exacerbated for subjects of information if there is a cultural disconnect between those evaluating the claim (the sources of evidence) and the subject of information. If there is a cultural disconnect, those evaluating the content may fail to appropriately understand it as well as its implications or implied sentiment, as these may all vary contextually across cultures. For example, social stereotypes affect how people understand language directed toward minority groups, leading to higher false negative assessments when labelers have negative stereotypes about the subject of information~\cite{davani2021hate}. One direction for future research would be to evaluate the impact of cultural alignment between the subject of (mis)information and the source of evidence to see if there are significant differences between a culturally-aligned and culturally-misaligned subject-source pairs. If one were to find differences between the subject-source pairs, this might imply that culturally-misaligned annotators provide lower signal (or high bias) labels.

\subsection{Evidence Retrieval}

\subsubsection{Machine Learning Task}

The next step in the pipeline involves retrieving evidence to either support or reject the claims that are deemed ``check-worthy''. Two main types of approaches have been proposed for this task. First, some methods (such as FAKTA~\cite{FAKTA}) utilize Google's commercial API to automatically search claims and retrieve (potentially) relevant evidence from the broader internet corpus. In some instances, this is followed by a post-processing steps in which results retrieved from search engines are merged with a database measuring the credibility of sources of information, and only evidence from sources deemed credible are kept~\cite{FAKTA}. 
The second type of approach is especially tailored to curb the spread of misinformation that has been previously fact-checked by humans. To do so, it uses a corpus of previously fact-checked claims stemming from either crowd-sourced knowledge, such as Wikipedia~\cite{wiki_evidence} or surveys~\cite{scaling_fact_checking}, or from professional assessments, such as Snopes or PolitiFact~\cite{politifact_evidence}. The task is then to map claims to previously fact-checked statements. 

In general, we can say that given a claim $X$ and a corpus $C$, each of which may have associated metadata ($X_{meta}$ and $C_{meta}$), the task is to retrieve a ranked list of the documents in the corpus $\overline{D}_{x}$, each with an associated relevance score $\overline{S}_{x}$, as shown in Equation~\ref{eq:f2}. 

\begin{equation}
    (\overline{D}_{x}, \overline{S}_{x}) = f_{2}(X, X_{meta}, C, C_{meta}) 
    \label{eq:f2}
\end{equation}

\subsubsection{Issues of Justice}

Typically, information retrieval algorithms are useful if they have both high recall and precision, as these are indicators of relevance and completeness of retrieved information, respectively \cite{manning2008introduction}. If returned evidence is disproportionately incomplete or irrelevant for some groups, then issues of justice may arise. Below, we posit several ways stakeholders might face issues of justice in automated evidence retrieval.

As previously noted, since the broader internet contains many documents making claims that have not been verified themselves, system designers may consider implementing an intermediate step to improve reliability of returned evidence: constraining the corpus of potential evidence to ``trusted'' sources, such as Wikipedia, PolitiFact, or other verified bodies of knowledge. However, these trusted sources of evidence may not engage with or have knowledge of claims from all peoples, and could exhibit bias in which claims they have chosen to fact-check. This could result in irrelevant and/or incomplete evidence gathered at this stage in the pipeline. Measures of recall could thus be relevant for assessing harms to seekers of information and sources of evidence. Seekers of information could be harmed if the corpus is limited to trusted sources of evidence that do not attend to some groups and communities, yielding low recall of information relevant to them. For sources of evidence, this may constitute an epistemic injustice, as failing to include them in the corpus of ``credible sources" unjustly diminishes its value and disregards the knowledge they have produced. 

Harms may also stem from the use of commercial search engines to retrieve evidence (with or without the intermediate step previously mentioned). Commercial search engines may down-weight evidence from websites with little traffic or poor search engine optimization (SEO), yielding different rankings for sources with the same relevance \(S_{x}\) as a result of maximizing for another commercial metric (click-through rate, time spent per link). If websites containing relevant information to support or reject claims from marginalized communities appear lower in search results despite having the same relevancy, this may inadvertently reduce the relevancy or completeness for the returned corpus of evidence \((C)\), which would again be reflected in recall metrics. This will constitute an epistemic injustice against sources of evidence by dismissing or discounting the information they produce in a way that does not hold an appropriate relationship to the quality of the information. Further, these issues may be exacerbated if only a relatively small corpus of potential evidence is kept after the claim is queried (e.g. first page of results) before document ranking and relevancy to the claim are determined through other means. Conversely, if websites that promote misleading evidence are ranked higher due to higher predicted engagement (i.e. ``click-bait'') rather than their relevance to the claim, this could render low precision in the evidence retrieval system. Such low precision metrics in relation to different types of information could potentially harm seekers and subjects of information; seekers could by affected through degraded fact-checking quality, while subjects could suffer recognitional injustices if false statements about them are encoded as part of a corpus treated as veridical.

\subsubsection{Sources of Harms}

Concerns over disparities in evidence retrieval are primarily grounded in properties of the current fact-checking ecosystem~\cite{duke_reporters_lab} and dynamics of the way information is gathered and disseminated on the web \cite{web_bias}. These may bias the composition of the corpus $C$, metadata regarding the corpus $C_{meta}$, or the proxies used to estimate $(\overline{D}_{x}, \overline{S}_{x})$. 

\begin{table}[ht]
\centering
\resizebox{\textwidth}{!}{\begin{tabular}{lrrr}
  \hline
Continent & Fact-Checking Organizations & Population (Billions) & Fact-Checking Organizations Per Billion \\ 
  \hline
Africa & 34 & 1.37 & 24.76 \\ 
Asia & 87 & 4.68 & 18.59 \\ 
Australia & 5 & 0.04 & 119.05 \\ 
Europe & 99 & 0.75 & 132.53 \\ 
North America & 81 & 0.60 & 135.91 \\ 
South America & 39 & 0.43 & 89.86 \\ 
   \hline
\end{tabular}}
\caption{This table shows the number of professional fact-checking organizations per billion population across 6 continents. The concentration of fact-checking organizations is roughly 7.3x higher in North America than in Asia when population is taken into account. (Source: \cite{duke_reporters_lab}) }
\label{tab:1}
\end{table}

While professional fact-checkers tend to deliver gold-standard evidence, they are significantly concentrated in North America and Europe, especially when accounting for population \cite{duke_reporters_lab}, as shown in Table \ref{tab:1}. Therefore, due to \textit{historical bias} in the data-generating process, there may be less previously fact-checked evidence available on the web for claims relevant to people from Asia, South America, and Africa. If claims regarding non-Western subjects of information \textit{are} fact-checked by a Western fact-checking organization, it will be examined from the perspective of Western experience and with a Western analytical framework \cite{journalism_western_bias}; this may induce a \textit{measurement bias} related to the stance of the fact-checking organization, and the contextual information they may or may not have access to. For instance, consider the stance a highly reputable news agency, \textit{The New York Times}, took towards the claim that Iraq possessed weapons of mass destruction. The newspaper famously published information stating that Iraq was in possession of weapons of mass destruction, only to, some time later, acknowledge their lack of integrity when interviewing sources of information. They said that their reporting was ''eagerly confirmed by United States officials convinced of the need to intervene in Iraq.'' \cite{editors_2004}

Another highly regarded source of evidence that is widely incorporated into fact-checking systems is Wikipedia~\cite{wiki_evidence}. While the premise of Wikipedia as an open-source encyclopedic body may yield hopes of a lower bias environment that is collectively constructed, in an early version of English Wikipedia, only 0.04\% of authors were responsible for the development (i.e. creation and related research) of over half the body of knowledge~\cite{web_bias}. Thus, ``the notion that it represents the wisdom of the overall crowd is an illusion'' [p. 56]. This effect has been called an \textit{activity bias}, and can impact the objectivity of open-source evidence if relevant groups do not contribute to the production of evidence. Furthermore, the amount of information available varies by subject and language. For instance, in a study comparing Polish and American Wikipedia articles about famous people, researchers noted that, ''English language entries have more references and external links... They also tend to be longer than Polish language entries''~\cite{wiki_bias_polish}.  

When search engines are used to retrieve information, \textit{learning bias} affecting the information retrieval algorithms may in turn bias the evidence retrieval system in undesirable ways. For example, if the commercial search engine used to retrieve the documents has the goal of optimizing revenue, using this as a ranking of relevance could have detrimental effects that could disproportionately affect sources of evidence, as well as seekers and subjects of information.

\textit{Measurement bias} may also emerge when relying on crowdsourced indicators of credibility. For example, FAKTA \cite{FAKTA} incorporates metadata \((C_{meta})\) regarding the credibility of sources of evidence, which is ultimately used to algorithmically weight the stance of evidence towards a claim. The metadata is pulled from an existing, crowd-sourced database\footnote{www.mediabiasfactcheck.com} that contains ratings on roughly 4,300 sources of evidence. If credibility is determined by a majority group, sources of evidence that are reliable but unknown or mis-assessed by the majority group will have their evidence down-weighted, even if the evidence is truthful and relevant to the claim. Future work should assess whether databases like this contain cultural biases in their assessment and composition of sources of evidence, which could bias \((\overline{D}_{x}, \overline{S}_{x})\).

\subsection{Automated Verdict Generation}

\subsubsection{Machine Learning Task}

After a claim has been determined to be ``check-worthy'' and relevant evidence has been gathered about the claim, the final phase involves generating a verdict regarding the claim. This is the most visible and consequential phase in the pipeline, as there is likely to be some significant action taken by the platform if a verdict is reached that the claim is false or misleading. Such systems may aim to provide labels visible to seekers of information, to assist with content moderation, or to provide decision support to human fact-checkers.

An important subtask that some have found relevant to generating a verdict is predicting the stance of each piece of relevant evidence towards the claim. In the standard version of this task, a document from the corpus \(C\) is probabilistically labeled (e.g. Supports, Against, Neither), depending on how it relates to the semantic representation of the claim \(X\). For more information on this task, see \citet{stance_detection_survey}. The predicted stance can then be viewed as metadata in relation to each evidence document, serving as input to a verdict generation algorithm. For example, FAKTA's end-to-end automated fact-detection system uses the stance and credibility of supporting pieces of evidence to yield a final prediction for the veracity of a claim \cite{FAKTA}. 

Other approaches have considered estimating the verdict of a claim without the need to retrieve supporting evidence. For example, \citet{supervised_fake_news_detection} construct semantic features extracted from the claim meant to capture subjectivity of language, features about the publisher of the claim meant to capture possible bias, as well as social media metadata surrounding the claim such as the number of likes and sharing patterns. 

In a generalization of \citet{evidence_aware_model}, the machine learning task can be formulated as follows:

\begin{equation}
     \overline{Y}_{v} = f_{3}(X, X_{meta}, C, C_{meta})
\end{equation}

In this equation, the verdict \(\overline{Y}_{v}\) is functionally related to the the semantic representation of claim \(X\), metadata of the claim \(X_{meta}\), a corpus of evidence \(C\), and the metadata regarding the corpus \(C_{meta}\), such as credibility and the predicted rank, relevance, and stance of each piece of evidence in relation to the claim. Naturally and as in the previous formulations, not all systems need to consider all inputs. The training data labels $\overline{Y}_{v}$ could be collected from professional fact-checkers, such as Snopes or PolitiFact, or they could be crowd-sourced as well, as in \citet{biases_verdict_perception} and \citet{scaling_fact_checking}. If there is more than one observation per claim, the labels from multiple crowd-sourced or professional fact-checkers are aggregated in some way (e.g. the mean, median, or ideological difference) to create a composite ``ground truth'' label \cite{biases_verdict_perception}.

\subsubsection{Issues of Justice}

Injustices to stakeholders occur when the predicted label \(\widehat{\overline{Y}}_{v}\) conflicts with the true, and potentially unobserved, label \(\overline{Y}_{v}\). In the case of a \textit{false negative}, in which no action (like labeling) is performed on potentially misleading information, significant harms are likely to be felt by seekers and subjects of information. As mentioned in Section \ref{sec:rel}, harms to seekers of misinformation are vast, and constitute broad threats to one's health, happiness, and societal institutions, especially if the misinformation is adversarial in nature (i.e. disinformation). Therefore, groups of seekers exposed to low quality information due to incorrect predictions face a distributive injustice of high quality of content moderation. Subjects of information, likewise, suffer recognitional injustice if their reputation is harmed by damaging and misleading information shared without a warning label, as seekers may interpret the claim to be true. This harm can be especially great to individual subjects of information in the court of public opinion. For instance, subjects of rumors spread on WhatsApp had their reputation damaged to such an extent that locals incorrectly thought they posed a grave threat, formed a mob, and murdered them \cite{whatsapp}.

A \textit{false positive} occurs when a moderation action is incorrectly taken against a claim made on a platform. When false positives are concentrated within some group of sources of information conditioned upon a particular sensitive attribute, this constitutes a participatory injustice. This is because the sources cannot contribute to or participate in the broader informational exchange at the same rate as others. If the action taken against the claim is to place a warning label on it, research has shown that this will reduce the credibility of the content \cite{labelling_cognitive}. Therefore, their credibility and capacity as a participant in the informational exchange has been unjustly diminished. Additionally, we can assume that a high rate of false positives will yield some reputational harm to the source of information, as their reputation might be damaged amongst their peers for sharing content that is routinely labeled as misleading. This harm would only be magnified if the source of information is also represented either expressly or contextually as the subject of information. For example, in the context of hate speech detection, a higher rate of false positives for posts in which a source of information discusses their identity of belonging to a disadvantaged group ~\cite{dixon2018measuring} affects stakeholders in their roles as both sources and subjects. Additionally, similar in nature to the harms associated with false negatives, seekers of information may be damaged when an action is incorrectly taken against a claim, especially if the action in consideration involves removing the claim from the platform. Seekers have a right, fundamentally, to view legitimate information unhindered, and this could qualify as a distributive injustice to access to legitimate information.

\subsubsection{Sources of Harms}

Research related to biases in determining the veracity of claims has covered two main topics, both related to the ``wisdom of crowds'' in determining misinformation from crowdsourcing labels. It is worth noting that existing research in this space has only assessed bias in the labels themselves, often considering crowdworkers as potential fact-checkers rather than as sources of training data for algorithms. Ways in which these biases affect algorithmic predictions are yet to be studied in this domain.  

One question that has been examined is whether obtaining labels from laypeople induces a \emph{measurement bias} when determining what constitutes misinformation. \citet{scaling_fact_checking} find that, while individual laypeople do tend to show considerably more bias and inconsistency than experts, a relatively small, politically balanced crowd yields an average label that is highly correlated to the average label of a few professional fact-checkers. This seems to imply that, under the stated circumstances, the measurement bias in the training labels is mitigated if the set of labelers is balanced with respect to a sensitive attribute $A=$\textit{political affiliation}. This finding suggests that there is wisdom in crowds, and provides a criteria for selecting the composition of layperson labeling teams. However, this research focuses largely on American political claims, so it is unclear whether a crowd of politically balanced laypeople maintain unbiased labels when the subject of information is non-Western or relatively unknown. 

Another important question to address is whether \emph{aggregation bias} exists when crowdsourced labels are averaged. In other words, does the crowd provide more signal beyond an ``average perception of truth''? By comparing crowdsourced perceptions to labels generated from professional fact-checking websites (such as Snopes), researchers have proposed several new signal-rich measures of ''perceptual difference'' amongst the crowd \cite{biases_verdict_perception}. For instance, they look at the marginal distributions of crowdsourced truth perceptions by political affiliation, which they refer to as ``ideological mean perception bias'' or IMPB. This is a promising step in assessing how differences between people affect their perceptions of truth in news. However, there may be differences in perceptions that extend beyond political ideology. Crucial to our study, this metric may be a useful way to compare perceptions of truth from groups of any protected attribute (race, gender, ethnicity, etc) to determine if there is a high level of disagreement between members of a particular group and professional fact checkers. If there are significant differences, the next logical step is to ask why this is the case. Does this group commonly encounter misinformation that has altered their perception? Or does this group have access to unique knowledge that gives them more expertise on a given topic than a general fact-checker? Some labelers may have epistemically advantaged standpoints~\cite{fazelpour2021diversity}, and aggregation mechanisms such as majority voting may fail to represent their assessments~\cite{davani2021dealing}.

\section{Conclusion}
In this paper, we employed and extended the informational justice framework to appropriately consider justice in algorithmic misinformation detection from the perspective of four key stakeholders: \textit{seekers of information}, \textit{sources of information}, \textit{subjects of information}, and \textit{sources of evidence}. Grounded in notions of \textit{distributive}, \textit{participatory}, \textit{recognitional}, and \textit{epistemic} justice, we analyzed the harms that might impact various stakeholders if algorithmic errors fall disproportionately on any group with a shared sensitive attribute. Specifically, we conducted this analysis across the three (generalized) stages of the automated pipeline: (1) \textit{check-worthiness}; (2) \textit{evidence retrieval}; and (3) \textit{verdict prediction}. We also put forth concrete ways to measure these harms in automated systems by drawing connections to widely used measures of algorithmic bias, and dissected potential sources of these biases.

In this work, the focus of our analysis has been the algorithmic prediction stages. However, downstream effects and dynamic human-algorithm interactions may yield novel concerns. For instance, a growing body of work has analyzed the effects of different interventions meant to curb misinformation~\cite{moravec2018fake, misinformation_media_literacy, badrinathan2021educative}. Thus, considering algorithmic errors in conjunction with the psychological effects of different interventions may be of importance, especially as it concerns the design and deployment of explainable fact-checking systems. 

A crucial challenge for algorithmic auditing of misinformation detection algorithms are the difficulties defining and labeling sensitive groups for the different stakeholders. In some cases, sources of information and/or evidence may have clear identities of belonging, e.g. when considering news media, Univision can be identified as Latinx media and Al Jazeera as a Middle Eastern source. Similarly, subjects of information can often be identified by leveraging computational linguistics to identify the subject of a claim when a group is explicitly mentioned in the claim. However, in some cases this is not straightforward and may have important ethical concerns associated with it~\cite{ghosh2021fair}. For example, attempting to infer gender can result in harms to transgender individuals~\cite{Hamidi.2018}.

In information ecosystems, justice is a multidimensional notion. The contributions presented in this paper provide a coherent framework for designers, policy makers and researchers to precisely articulate and identify the varied ways in which concerns about injustice can arise in algorithmic misinformation detection. In doing so, it offers guidance to efforts towards developing technical responses that counter the threat of misinformation in an equitable manner.

\section*{Funding Acknowledgment}
This work was supported in part by Good Systems, a research grand challenge at the University of Texas at Austin, and by a Google AI Award for Inclusion Research. 

\bibliographystyle{ACM-Reference-Format}
\bibliography{sample}

\end{document}